\documentclass{ws-procs975x65}            

\def\beq{\begin{equation}}
\def\eeq{\end{equation}}
\def\bea{\begin{eqnarray}}
\def\eea{\end{eqnarray}}
\def\bealn{\begin{eqnarray}}
\def\eealn{\end{eqnarray}}

\begin{document}
\title{\bf The Jet Energy Profile:\\
A BSM Analysis Tool\footnote{This contribution is an abbreviated version of ref. \cite{Chivukula:2014pma}.}}

\author{R. S. Chivukula$^\#$\footnote{Dr. Chivukula presented this talk at SCGT15.}, E. H. Simmons, and N. Vignaroli}

\address{Department of Physics and Astronomy, Michigan State University,\\
East Lansing, MI, 48825, USA\\
$^\#$E-mail: sekhar@msu.edu}

\begin{abstract}
A new heavy di-jet resonance could be discovered at the 14 TeV LHC. In this talk we present a strategy to reveal the nature of such a particle; in particular to discern whether it is a quark-antiquark ($q\bar{q}$), quark-gluon ($qg$), or gluon-gluon ($gg$) resonance. The strategy is based on the study of the energy profiles of the two leading jets in the di-jet channel. Including statistical uncertainties in the signal and the QCD backgrounds, we show that one can distinguish between $gg$, $qg$, and $q\bar{q}$ resonances; an evaluation of systematic uncertainties in the measurement of the jet energy profile will require a detailed detector study once sufficient 14 TeV di-jet data is in hand.
\end{abstract}


\bodymatter

\section{Introduction}
Searches for heavy resonances produced in the $s$-channel and decaying into a pair of jets offer a simple and powerful probe of many different scenarios of new physics at the Large Hadron Collider. ATLAS \cite{Aad:2014aqa} and CMS \cite{CMS:kxa, Chatrchyan:2013qha} have recently presented the results of the searches for narrow di-jet resonances at the LHC with $\sqrt{s}=8$ TeV. Lower limits on the masses of new hypothetical particles in a variety of beyond the standard model theories have been obtained. The upcoming LHC run at $\sqrt{s}=14$ TeV will have the capability to greatly extend the discovery reach in the di-jet channel \cite{Gershtein:2013iqa}. If a hadronic resonance is discovered in the di-jet channel a major challenge will be the identification of the nature and of the properties of the newly discovered particle. In this talk we present a strategy to reveal if such a particle is a quark-antiquark ($\bar{q}q$), a quark-gluon ($qg$), or a gluon-gluon ($gg$) resonance. This strategy\cite{Chivukula:2014pma} analyzes the energy profiles of the two final jets \cite{Ellis:1992qq}.     

We consider three compelling benchmark scenarios to describe different di-jet resonances: the flavor universal coloron model \cite{Simmons:1996fz, Chivukula:1996yr} for $\bar{q}q$ resonances, the excited quark model of Ref. \cite{Baur:1989kv, Baur:1987ga} for $qg$ resonances and the general parameterization in \cite{Han:2010rf} of color-octet scalar interactions for $gg$ resonances. 
All of the results are shown in the relevant mass-coupling parameter space that is both not excluded by the 8 TeV LHC analyses \cite{Aad:2014aqa, CMS:kxa, Chatrchyan:2013qha} and conducive to a 5$\sigma$ discovery of the resonance in the di-jet channel at the 14 TeV LHC. The LHC-8 excluded regions are extracted from the ATLAS \cite{Aad:2014aqa} and CMS \cite{CMS:kxa, Chatrchyan:2013qha} searches; the LHC-14 discovery reach is evaluated based on Monte Carlo simulations. 

Information on the partonic origin of a di-jet resonance is provided by the analysis of the jets' substructure, in particular by the study of the energy profiles \cite{Ellis:1992qq} of the two final jets. Quark-initiated jets have more quickly rising profiles compared to gluon jets, so that discrimination among the different $\bar{q}q$, $qg$, and $gg$ di-jet resonances is possible from analyzing the di-jet energy profiles. We evaluate the (mean) jet energy profiles by applying the theoretical calculations in perturbative QCD, at next-to-leading logarithmic accuracy, which have been developed in Ref. \cite{Li:2011hy,Li:2012bw}. Statistical fluctuations on the jet energy profiles are generated through Monte Carlo simulations and, consequently, the statistical efficiency of our discriminating tool based on the di-jet energy profiles is evaluated. Including statistical uncertainties in the signal and the background, we show that one can distinguish between $gg$, $qg$, and $q\bar{q}$ resonances; an evaluation of systematic uncertainties in the measurement of the jet energy profile will require a detailed detector study once sufficient 14 TeV di-jet data is in hand and is beyond the scope of this paper. \footnote{Additional techniques based on the study of jet substructure and aimed at identifying di-jet resonances and/or improving the signal-to-background ratio have been extensively considered in the literature. Recent examples are the study of the color flow in Ref. \cite{Gallicchio:2010sw}, the analysis of the charge track multiplicity and the $p_T$-weighted linear radial moment (girth) in Refs. \cite{Gallicchio:2011xc, Gallicchio:2011xq}, and the study of generalized angularities in \cite{Larkoski:2014pca} aimed at distinguishing quark and gluon jets on an event-by-event basis. The current status of jet substructure techniques, covering both experimental and theoretical efforts, is reviewed in \cite{Altheimer:2012mn}. More sophisticated analyses employing these methods could yield even more discrimination power than the simpler method considered here.}

\section{Benchmark models for di-jet resonances}\label{sec:models}

We consider three benchmark models for $\bar{q}q$, $qg$ and $gg$ resonances. For $gg$ and $qg$ resonances, we will refer to the same models as were considered in the recent CMS \cite{CMS:kxa, Chatrchyan:2013qha} and ATLAS \cite{Aad:2014aqa} analyses.   For the $\bar{q}q$ resonance, we will consider the flavor universal coloron model considered in the CMS analysis \cite{CMS:kxa, Chatrchyan:2013qha}.

\subsection{Flavor universal colorons ($\mathbf{C}$)}

Quark-antiquark resonances are present in many different kinds of new physics scenarios. 
We focus here  on the coloron model 
presented in Ref. \cite{Simmons:1996fz, Chivukula:1996yr}.  This model belongs to the class of theories predicting an extended strongly interacting sector $SU(3)_1 \times SU(3)_2$ that spontaneously breaks to $SU(3)_{QCD}$ 
\cite{Hill:1991at, Frampton:1987dn, Martynov:2009en}. The model can be flavor universal, which is the case we will consider here.

At high energies, the model features an enlarged color gauge structure $SU(3)_1 \times SU(3)_2$. This extended color symmetry is broken down to $SU(3)_C$ by the (diagonal) expectation value of a scalar field, which transforms as a ({$\bf 3, \bar{3} $}) under $SU(3)_1 \times SU(3)_2$. 
It is assumed that each standard model (SM) quark transforms as a $\bf (1,3)$ under the extended strong gauge group.
The color symmetry breaking induces a mixing between the original $SU(3)_1$ and $SU(3)_2$ gauge fields, which is diagonalized by a field rotation determined by
\begin{equation}\label{eq:tgtheta}
\tan\theta=\frac{g_2}{g_1} \qquad g_S = g_1 \sin\theta = g_2 \cos\theta \ ,
\end{equation} 
where $g_S$ is the QCD strong coupling and $g_1$, $g_2$ are the $SU(3)_1$ and $SU(3)_2$ gauge couplings, respectively. The diagonalization reveals two classes of color-octet vector boson mass eigenstates --  the massless SM gluons and the new colorons $C^{a}$, which are massive,
\begin{equation}\label{eq:c-mass}
 m_{C}=\frac{g_S u}{\sin\theta \cos\theta} \ ,
\end{equation} 
where $u$ is the breaking scale for the extended color symmetry. The coloron's interactions with quarks are determined by a new QCD-like coupling
\begin{equation}\label{eq:color-current}
- g_S \tan\theta \sum_f \bar{q}_f \gamma^{\mu} \frac{\lambda^{a}}{2} q_f C^{a}_{\mu} \ .
\end{equation}   
A coloron that decays to all six quark flavors ($m_c > 2m_{top}$) has a decay width:
\begin{equation}\label{eq:Col-Gamma}
\Gamma(C)=\alpha_{S}  m_C \tan^2{\theta}
\end{equation}

Colorons can be produced at the LHC by quark-antiquark fusion at a rate determined by the $C$ coupling to light quarks, $g_s \tan\theta$.  Gluon-gluon fusion production, on the other hand, is forbidden at tree level by $SU(3)_C$ gauge invariance   \cite{Cho:1995vh, Zerwekh:2001uq, Chivukula:2001gv}, and has been found to be insignificant at the one-loop level \cite{Chivukula:2013xla}.  The CMS search for di-jet resonances \cite{CMS:kxa, Chatrchyan:2013qha} has considered the hypothesis of a flavor universal coloron, taking this model as a benchmark and fixing $\tan\theta=1$.

\subsection{Excited quarks ($\mathbf{q^*}$)}

Quark-gluon resonances are a general prediction of composite models with excited quarks \cite{Baur:1989kv, Baur:1987ga}. In this work we will take as our exemplar the phenomenological model of \cite{Baur:1989kv}, which describes an electroweak doublet of excited color-triplet vector-like quarks $q^{*}=(u^{*}, d^{*})$ coupled to first-generation ordinary quarks. In this model, 
right-handed excited  quarks interact with gauge bosons and ordinary (left-handed) quarks through magnetic moment interactions described by  the effective Lagrangian:
 \begin{align}
 \begin{split}\label{eq:L_qStar}
& \mathcal{L}_{int}=\frac{1}{2\Lambda} \bar{q}^{*}_R \sigma^{\mu\nu} \left[ g_S f_S \frac{\lambda^{a}}{2} G^{a}_{\mu\nu}+g f \frac{\mathbf{\tau}}{2}\cdot \mathbf{W}_{\mu\nu} + g^{'} f^{'} \frac{Y}{2} B_{\mu\nu} \right]  q_L + \text{H.c}. 
 \end{split}
 \end{align}

The excited quarks can decay into $qg$ or into a quark plus a gauge boson. The corresponding decay rates are:
 \begin{align}
 \begin{split}\label{eq:Gamma_qStar}
 &\Gamma(q^{*}\to qg)=\frac{1}{3}\alpha_S f^2_{S}\frac{m^3_{q*}}{\Lambda^2} \\ 
 &\Gamma(q^{*}\to q\gamma)=\frac{1}{4}\alpha f^2_{\gamma}\frac{m^3_{q*}}{\Lambda^2} \\ 
 &\Gamma(q^{*}\to qV))=\frac{1}{8}\frac{g^2_V}{4\pi}f^2_{V}\frac{m^3_{q*}}{\Lambda^2}\left[ 1-\frac{m^2_V}{m^2_{q*}}\right]^2\left[ 2+\frac{m^2_V}{m^2_{q*}}\right]
 \end{split}
 \end{align}
 with $V=W,Z$ and with the definitions

\begin{align}
 \begin{split}\label{eq:f_qStar}
 & f_{\gamma}=f T_3 +f'\frac{Y}{2}\\
  & f_{Z}=f T_3\cos^2\theta_W -f'\frac{Y}{2}\sin^2\theta_W\\
  & f_{W}=\frac{f}{\sqrt{2}}~.
 \end{split}
 \end{align}
The $q^{*}\to qg$ branching ratio is about 0.8 for $f_S=f=f'$.

Excited quarks are singly produced at the LHC through quark-gluon annihilation and, as just noted, they dominantly decay into $qg$. For our analysis, we choose the benchmark parameters $\Lambda=m_{q^{*}}$ and $f_S=f=f'$, while allowing the overall coupling strength to vary. By way of comparison, recent LHC searches, CMS \cite{CMS:kxa, Chatrchyan:2013qha} and ATLAS \cite{Aad:2014aqa} have used the same value of $\Lambda$ with $f_S=f=f'=1$.

\subsection{Color-octet scalars ($\mathbf{S_8}$)}

A gluon-gluon final state can generally arise from decay of colored scalars in models with extended color gauge structures \cite{Hill:1991at, Frampton:1987dn, Martynov:2009en, Chivukula:2013hga, Bai:2010dj}. 
In this work we adopt the general effective interaction for a color octet scalar, $S_8$, introduced in \cite{Han:2010rf}:
\begin{equation}\label{eq:L-s8}
\mathcal{L}_{S_8}=g_S d^{ABC}\frac{k_S}{\Lambda_S} S^A_8 G^B_{\mu\nu}G^{C, \mu\nu} \ ,
\end{equation}
where $d$ is the QCD totally symmetric tensor.

A colored scalar of this kind is singly-produced at the LHC through gluon-gluon annihilation. We consider the case in which it decays entirely (or almost entirely) into gluons. The corresponding decay rate reads:
\begin{equation}\label{eq:Gamma-s8}
\Gamma(S_8)=\frac{5}{3}\alpha_S \frac{k^2_S}{\Lambda^2_S}m^3_{S_8}~.
\end{equation}
We set $\Lambda_S=m_{S_8}$ and we present results for different couplings $k_S$.  Similarly, CMS \cite{CMS:kxa, Chatrchyan:2013qha} and ATLAS \cite{Aad:2014aqa} present searches for $\Lambda_S=m_{S_8}$ and $k_S=1$.

\section{LHC Discovery Reach}\label{sec:reach}

For each type of dijet resonance, we begin by deriving the relevant mass and coupling parameter space for our analysis, namely the region that is not yet excluded by LHC-8 analyses and in which a 5$\sigma$ discovery will be possible at the 14 TeV LHC. 

We derive the excluded parameter region for colorons from the CMS analysis in \cite{CMS:kxa}; for excited quarks
and scalar-octets, we obtain constraints by considering the strongest limits within the CMS \cite{CMS:kxa} and ATLAS analyses \cite{Aad:2014aqa}. 
Note that CMS and ATLAS searches have a poor sensitivity to resonances whose width is large compared to the detector di-jet mass resolution, {\it i.e.} with a width-over-mass value of greater than $\sim$0.15  \cite{Harris:2011bh}.
 In what follows, we assume that the new resonances are sufficiently narrow to be discovered in di-jets and that they decay only (or at least predominantly) to pairs of jets: $\bar{q}q$, $qg$, or $gg$.

The 5$\sigma$ discovery reach at the 14 TeV LHC is estimated by evaluating $S/\sqrt{S+B}$, where $S$ and $B$ are, respectively, the total number of signal and background di-jet events passing the CMS kinematic selection criteria in \cite{CMS:kxa}:
\begin{equation}\label{eq:cms-cut}
p_T (j_{1,2}) > 30 \ \text{GeV} , \qquad |\eta (j_{1,2})| < 2.5 \ , \qquad |\Delta \eta (j_1 j_2)|<1.3~.
\end{equation}
For a given potential resonance mass $M$ we also require the invariant mass of the two leading jets to be within a range of $\pm 0.15 M$ from the di-jet invariant mass peak. The standard model di-jet background is taken from Ref. \cite{Yu:2013wta}, where it has been carefully estimated by applying the same CMS cuts to matched samples of two- and three-jet final states using MADGRAPH \cite{Alwall:2011uj} and PYTHIA \cite{Sjostrand:2007gs}. The simulated di-jet signals at the 14 TeV LHC for the different resonances are generated at parton level with MADGRAPHv5 and the CT10 \cite{Lai:2010vv} set of parton distribution functions, after implementing the benchmark models with Feynrules \cite{Christensen:2008py}. We find an acceptance rate for the CMS kinematic selection criteria  \cite{CMS:kxa} of about 0.58 for $S_8$ or $q^{*}$ and of about 0.5 for $C$, for the mass ranges of interest. 

\begin{figure}
\centering
\includegraphics[width=0.58\textwidth]{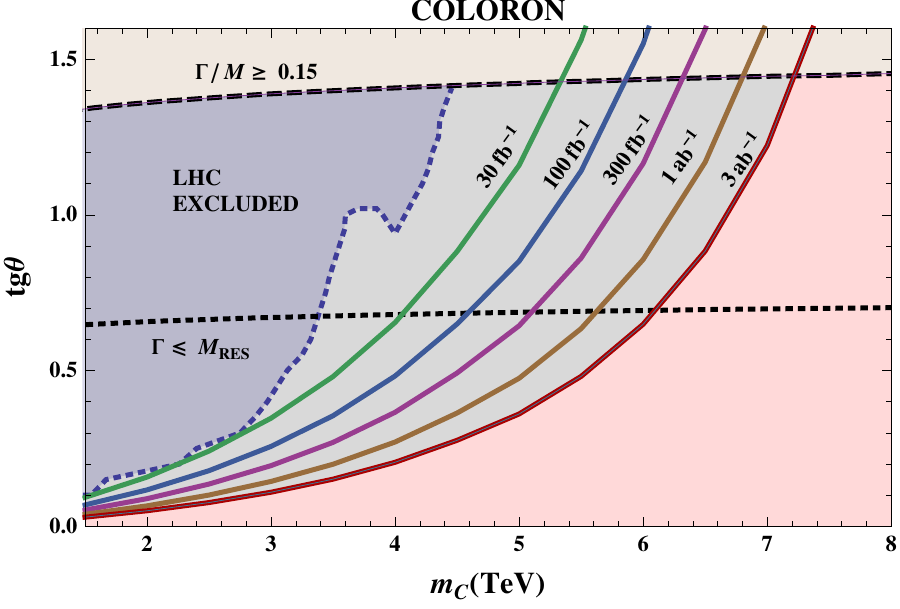}
\includegraphics[width=0.58\textwidth]{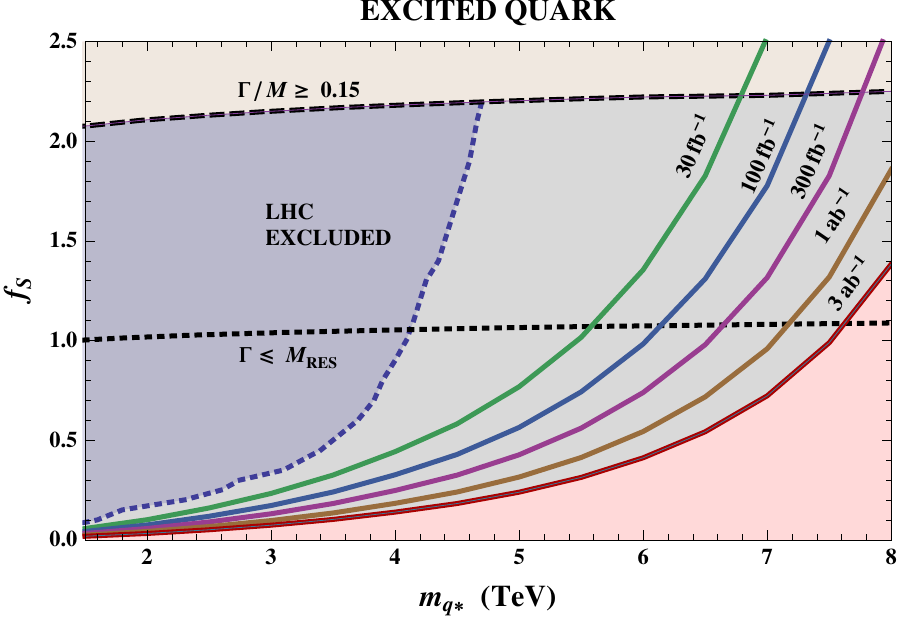}
\includegraphics[width=0.58\textwidth]{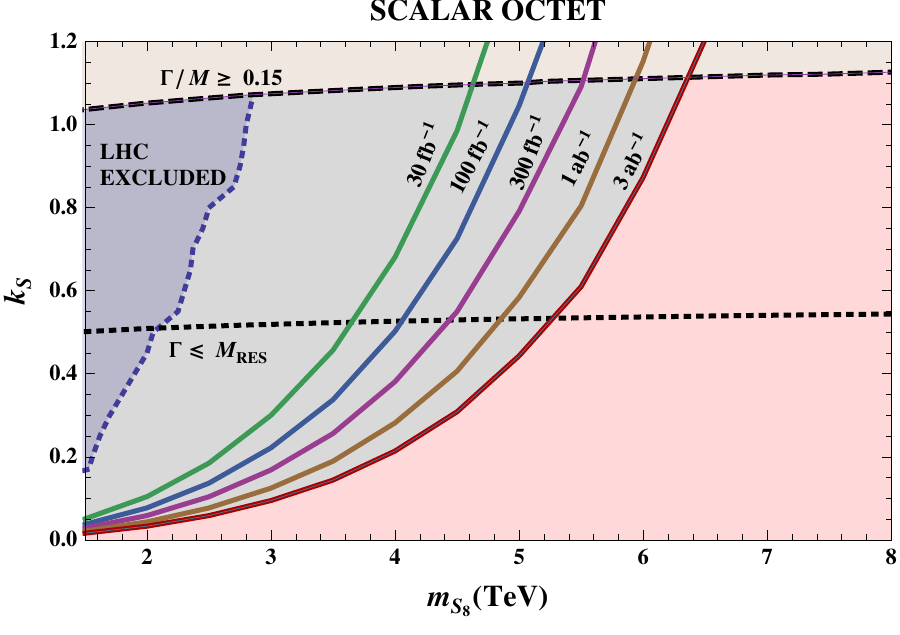}
\caption{\small In each pane from left to right: regions of coupling-mass parameter space excluded by LHC-8 (blue), regions accessible to LHC-14 (pale grey), region inaccessible at LHC-14 (pink). Thick colored curves show the 5$\sigma$ reach at luminosities from 30 -- 3000 fb$^{-1}$. Above the upper dashed line, the resonance is too broad to detect ($\Gamma=0.15 M$); below the lower dotted line it is narrower than the experimental resolution ($\Gamma\leq M_{res}$), where $M_{res} = 0.035 M$ \cite{CMS:kxa}. Resonance widths are calculated as shown in section \ref{sec:models}.  }
\label{fig:excl-disc}
\end{figure}

Fig. \ref{fig:excl-disc} shows our estimates of the 5$\sigma$ reach at the 14 TeV LHC in the mass-vs.-coupling plane for colorons, excited quarks, and scalar-octets, for integrated luminosities of 30 -- 3000 fb$^{-1}$. The discovery reach we find for the coloron is very similar to those already derived in \cite{Yu:2013wta, Dobrescu:2013cmh} and in \cite{Atre:2013mja}. 

Within each pane of Fig. \ref{fig:excl-disc}, we may identify a ``region of interest" where a resonance of a given mass and coupling is not excluded by LHC (i.e., is not in the blue region at left), is relatively narrow (lies below the horizontal dashed curve) and would be detectable at LHC-14 at the indicated luminosity (is within the central light-grey region).  
%

\section{Jet Energy Profiles}\label{sec:JEP}

In this section we will examine the use of jet energy profiles (JEPs) \cite{Ellis:1992qq} to statistically distinguish $\bar{q}q$, $qg$ and $gg$ di-jet resonances. 
For a jet of size $R$, the (integrated) JEP, $\psi(r)$, is defined as the fraction of jet transverse momentum that lies inside a sub-cone of size $r\ (<R)$,
\begin{equation}
\label{eq:psi}
\psi(r) = \frac{ \sum\limits_{r'<r}^{} p_T(r')}{\sum\limits_{r'<R}^{} p_T(r')}.
\end{equation}
Gluon-initiated jets radiate more and produce a slowly rising JEP. Quark initiated jets, on the other hand, radiate less and have a quickly rising JEP. 

We will begin by considering the statistical limitations of measuring the JEP of a simulated sample of {\it pure} signal events -- di-jet events arising solely from a coloron, excited quark, or color-octet scalar. We will subsequently consider the effect of QCD background on the statistical significance of measuring the jet energy profile of the signal. We will show that the measurement is not statistically limited and, if systematic errors can be controlled, will clearly distinguish between the types of di-jet resonances we consider. An evaluation of systematic uncertainties in the measurement of the jet energy profile will require a detailed detector study once sufficient 14 TeV di-jet data is in hand, and is beyond the scope of this paper.

\subsection{JEP Measurement Based on Signal Events}

We consider first the measurement of the jet energy profile of a sample of di-jet events arising solely from the production
of a coloron, excited quark, or color-octet scalar. As explained above, due to the differing pattern of soft gluon radiation from quarks and from gluons, in principle a measurement of the JEP could distinguish among 
these different types of resonances since they decay into different final states. Experimentally measured JEPs, of course, include not just the effects of the initial high-$Q^2$ radiation arising from the quarks and gluons produced in the hard event, but the subsequent low-$Q^2$ showering and hadronization of these objects -- a description of which depends on tune-dependent Monte Carlo event generators such as PYTHIA \cite{Sjostrand:2007gs,ATLAS:2012uec}.
JEPs have been recently measured at ATLAS \cite{Aad:2011kq} and at CMS \cite{Chatrchyan:2012mec} and, indeed,
the results of the JEP measurements \cite{Aad:2011kq,Chatrchyan:2012mec} show that the data can be reproduced only after a careful calibration of the shower/hadronization parameters.

The copious di-jet data available from the 14 TeV LHC will allow for the
necessary calibration -- and, as we will show, we expect to find clear differences between the di-jet JEP measured in the resonance region and that measured from the purely SM background events at off-resonance di-jet invariant masses. However, since 14 TeV LHC data and tuned event
generators are not yet available, we will rely on a theoretical calculation to estimate the {\it average} shape of the JEPs for colorons, excited quarks, and color-octet scalars, and we will use MC simulations -- MADGRAPH interfaced with PYTHIA v6.4 (default tune) -- 
to evaluate the statistical uncertainties on the measurement of these profiles.
Specifically, we calculate the mean values of the jet energy profiles in perturbative QCD (pQCD)  by using the jet functions derived in \cite{Li:2011hy,Li:2012bw}, which apply a next-to-leading-logarithm (NLL) resummation\footnote{Terms of the form $\alpha_S ^n (\log (R/r))^{2n}$, $\alpha_S ^n (\log (R/r))^{2n-2}$ are resummed to all orders in $\alpha_S$. The studies in \cite{Li:2011hy,Li:2012bw} show that NLL resummation is necessary for a correct description of the data; fixed NLO calculations overestimate the JEPs and fail to describe the data.}.  Indeed, we find this procedure yields very good agreement with the experimental data from CMS at 7 TeV \cite{Chatrchyan:2012mec} and the Tevatron \cite{Acosta:2005ix}. The pQCD calculation depends on two phenomenological parameters that take into account the effect of uncalculated sub-sub-leading logarithmic contributions. We will fix these parameters at the values that reproduce the Tevatron data \cite{Acosta:2005ix}. Once calibration becomes possible, these parameters, too, will need to be fixed at the values that reproduce the 14 TeV LHC data\footnote{{\it E.g.}, $Z$
+jets events with jets in a kinematic region similar to that of di-jet resonances could be used as calibration samples.}.  Since we are not using calculations tuned to LHC energies, our {\it absolute} results for the  jet energy profiles will not precisely match those to be expected at the LHC --- however, we expect the {\it relative differences} in the JEPs we find between the various kinds of resonances to be representative of what would be seen there.

We consider first the signal of a 4 TeV di-jet resonance, coming from an $S_8$, $C$ or $q^{*}$, which can be discovered with approximately 30 fb$^{-1}$ at the 14 TeV LHC and which has not been excluded by the present LHC-8 searches. In particular, we consider an $S_8$ resonance with a coupling $k_S=0.65$, a coloron with $\tan\theta=0.6$ and a $q^{*}$ with $f_S=0.4$. After the CMS selection cuts (\ref{eq:cms-cut}), all of these three types of resonances give, approximately, the same signal cross section around the resonance peak. We will analyze the jet-energy profiles for di-jet resonance events passing the CMS kinematic cuts (\ref{eq:cms-cut}) and in a region $|M_{jj}-M|<\Gamma/2$; we take conservatively $\Gamma=\Gamma(S_8)$, corresponding to the largest possible width among those of the three types of resonances. The choice of focusing our analysis in a narrow region around the resonance peak is intended to minimize  the SM di-jet background which, as we will see in the next subsection, will affect the uncertainty of our discriminating tool.  After selection we obtain a di-jet resonance signal cross-section of 22~fb. 

The predicted JEPs for a quark or gluon jet, $\psi(r)$, are obtained as in Ref. \cite{Rentala:2013uaa} by fitting a functional form to the results of a full perturbative QCD calculation done at several values of $r$.\footnote{To be more specific, the predicted $\psi(r)$ JEPs for either quark jets or gluon jets are obtained by fitting an exponential function of the type $(1-b e^{-ar})/(1-b e^{-aR})$ \cite{Rentala:2013uaa} to the discrete $\psi(r)$ values obtained from the full pQCD calculation at several fixed $r$ points. We calculate $\psi(r)$ at $\Delta r=0.1$ steps, starting from $r=0.1$ up to $r=R=0.5$.} Since the resonances we are studying each decay to two jets, we then calculate a predicted di-jet profile $\psi_{jj}$ for the resonance decay as 
\begin{equation}\label{eq:mean-JEP}
\psi_{jj}(r)=\psi_1(r)+\psi_2(r)
\end{equation}
where $\psi_1(r)$, $\psi_2(r)$, respectively, denote the JEPs of the leading and next-to-leading jet. 

In order to quantify the power of JEPs to discriminate between different types of resonances, we will apply a one-parameter fit, so that we can unequivocally assign a specific value of the fit parameter to each signal di-jet profile. Specifically, we can parameterize the generic di-jet profile of the signal as
\begin{equation}\label{eq:f-param}
\psi_S(r)=f \psi_{\bar{q}q}(r) + (1-f)\psi_{gg}(r) \ .
\end{equation} 
Here, $f$ is our fit parameter that indicates the fraction of quark-jets in a generic di-jet resonance: $f=0, 0.5, 1$ for a $gg$, $qg$, or $\bar{q}q$ resonance respectively. The mean values of the different jet-energy profiles determined by pQCD are shown as the central values of the curves in Fig. \ref{fig:JEP} -- note the difference between the JEPs arising from $\bar{q}q$, $qg$, and $gg$ dijet events.

We estimate the statistical uncertainty on the mean values of the JEPs for a sample of pure signal events by running pseudo-experiments through MC simulations. We evaluate the statistical errors in the JEPs at $\Delta r=0.1$ steps. Signal sample events are generated with MADGRAPH v.5 \cite{Alwall:2011uj} and interfaced with PYTHIA v.6 for shower and hadronization. 
The jets are clustered through FASTJET \cite{Cacciari:2011ma} by an anti-kT algorithm with cone size $R=0.5$. JEPs are then obtained by analyzing the jet substructure, according to the formula in (\ref{eq:psi}).
We find, as expected, that the statistical fluctuations in $\psi$, and hence $f$, follow Gaussian distributions and that the errors scale as the square root of the number of events. In particular, we find that
the uncertainty in the value of $\psi(r)$ at $r=0.1$ (which yields the largest error) scales as
\begin{equation}
(\delta \psi_S(0.1))^2 \approx \frac{\sigma^2(0.1)}{S}~,
\label{eq:psi-error}
\end{equation}
where $\sigma(0.1) \approx 0.4$ and $S$ is the total number of signal events.

\subsection{Including QCD Background}

\begin{figure}
\begin{center}
\includegraphics[width=0.6\textwidth]{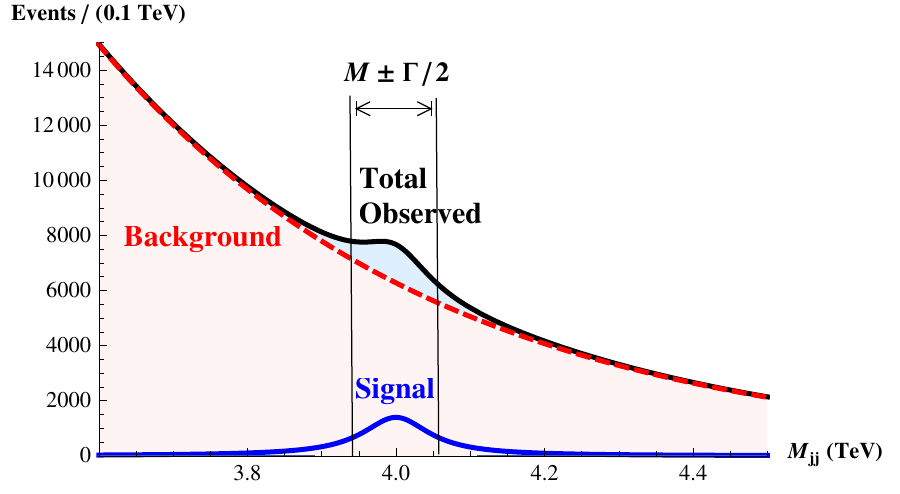}
\end{center}
\caption{ \small Sketch of how the number of signal (blue), background (red) and total observed events (black) could depend on dijet invariant mass ($M_{jj}$) if a dijet resonance is discovered at the LHC.  This figure illustrates issues raised in the discussion of JEP measurements and uncertainties in Section 5.2.}
\label{fig:side-band}
\end{figure}

Next, we consider the impact of QCD background on our analysis. The resonance will appear as a ``bump" in a plot of the di-jet invariant mass distribution, as sketched in Fig. \ref{fig:side-band}. In the signal region ($|M_{jj} - M| < \Gamma/2$) there will be $S$ signal events and $B$ QCD background events. As mentioned above, for the benchmark 4 TeV di-jet resonance we find a signal cross section of 22~fb, and extracting the background from Ref. \cite{Yu:2013wta}, we find a signal-to-background ratio of 1/23. 

It is not possible to measure the jet energy profile of the signal alone; measurements of the JEP in the signal region, $\psi_{OBS}(r)$, will include both signal and background. One can also measure the jet energy profiles in ``side-bands", regions of di-jet invariant mass immediately adjacent to but outside the resonance region; this yields an experimentally determined measurement of the JEP of the QCD background, $\psi_B(r)$. We expect that the experimental uncertainties on these individual measurements will scale analogously to what is shown in eq. (\ref{eq:psi-error}):
\begin{equation}
(\delta \psi_{OBS}(r))^2 \approx \frac{\sigma^2(r)}{S+B}\,\hskip20pt
(\delta \psi_{B}(r))^2 \approx \frac{\sigma^2(r)}{B}~.
\label{eq:morepsi-error}
\end{equation}

\begin{figure}
\centering
\includegraphics[width=0.6\textwidth]{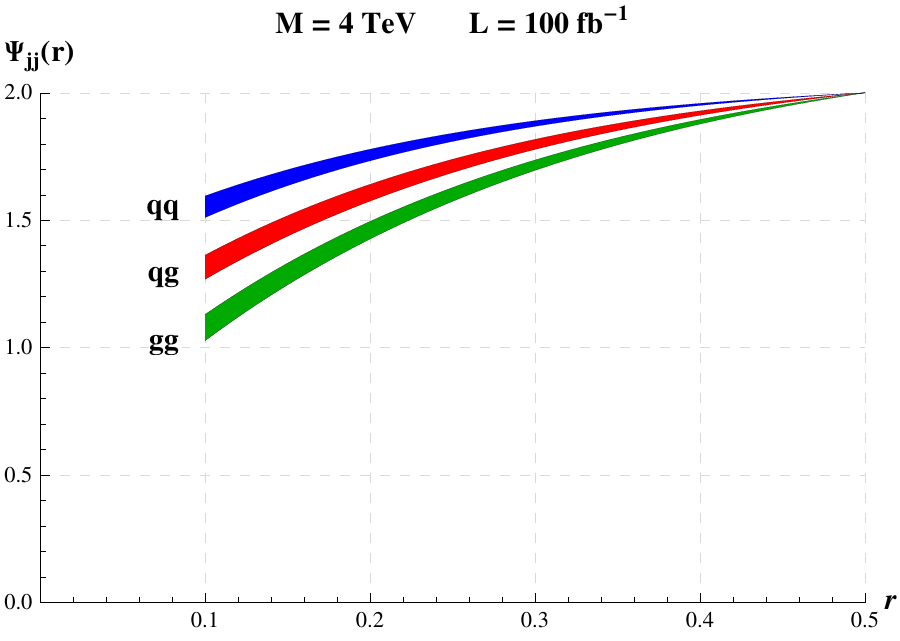}
\caption{\small Di-jet energy profiles for $\bar{q}q$ (coloron), $qg$ (excited quark) and $gg$ (scalar octet) 4 TeV di-jet resonances (the respective resonance couplings are fixed to $\tan\theta=0.6$, $f_{S} =0.4$, and $k_{S} =0.65$). Each band shows a $\pm 1\sigma$ statistical variation from the mean curve. The effect of background subtraction, eq. (\ref{eq:total-error}), is included. }
\label{fig:JEP}
\end{figure}

The desired quantity $\psi_S(r)$ is now related to the measurable JEPs by
\begin{equation}
\psi_{OBS}(r) = \frac{S}{S+B} \psi_S(r) + \frac{B}{S+B}\psi_B(r)~,
\end{equation}
and hence
\begin{equation}\label{eq:B-subtract}
\psi_S(r) = \psi_{OBS}(r) +\frac{B}{S}(\psi_{OBS}(r)-\psi_{B}(r))~.
   \end{equation}
The statistical uncertainties in the quantities $\psi_{OBS}$ and $\psi_{B}$ are given by eq. (\ref{eq:morepsi-error}).
Since we are working in a regime in which $B \gg S$, the uncertainty in $B/S$ is dominated by
fluctuations in the number of signal events and is roughly $B/S^{3/2}$. From eq. (\ref{eq:B-subtract}), 
we find the mean-square error on $\psi_S$ to be
\begin{equation}
(\delta \psi_S)^2 \approx \frac{\sigma^2}{S}\left[1+2\frac{B}{S}\right] + \frac{(\psi_S-\psi_B)^2}{S}~
\label{eq:total-error}
\end{equation}
where we have neglected terms suppressed by $S/B$. The first term in eq. (\ref{eq:total-error}) represents the
``dilution" in the measurement of $\psi_S$ due to QCD background, relative to the sample-only error of eq. (\ref{eq:psi-error}), and the second term is due to the uncertainty in the number of signal events. From Fig. \ref{fig:JEP}, we see that the difference in JEPs (which is maximal for the difference between pure $qq$ and $gg$ states at $r=0.1$) is bounded from above by about 0.5; in the regions in which the di-jet resonance can be observed, the second term in eq. (\ref{eq:total-error}) is negligible.
Fig. \ref{fig:JEP} shows the resulting di-jet energy profiles, with uncertainty bands including the effect of the background subtraction, for the $\bar{q}q$ (coloron), $qg$ (excited quark) and $gg$ (scalar octet) 4 TeV di-jet resonance.

We can translate the statistical error on $\psi(r)_S$ into a statistical uncertainty on the $f$ parameter \footnote{This is obtained via the following procedure. Using a step size $\Delta r=0.1$, we generate a large number of $\psi(r)$ values according to the Gaussian fluctuations which we have calculated by running pseudo-experiments. The generated $\psi(r)$ points are fitted by the function $(1-b e^{-ar})/(1-b e^{-aR})$ and the resulting profiles are translated into $f$ values according to eq. (\ref{eq:f-param}). We thus obtain the statistical fluctuation on $f$ and we are able to calculate the corresponding standard deviation.  }. Results predicted for the 14 TeV LHC with 100 fb$^{-1}$ of data are $f=1.00 \pm 0.06$ for a $q\bar{q}$ resonance, $0.50\pm 0.07$ for a $qg$ resonance, and $0.00 \pm 0.08$ for a $gg$ resonance.

We can also evaluate the statistical efficiency of our discriminating tool according to a $t$-test. We find that even the most challenging discrimination, that between $qg$ and $gg$ resonances, can be performed at a high statistical level, of $\sim$5$\sigma$, with 100 fb$^{-1}$ at the 14 TeV LHC. A $\bar{q}q$ resonance can be well distinguished from a $gg$ resonance: a $\sim$5$\sigma$ level of distinction can be achieved with only $\sim$30 fb$^{-1}$.

We can repeat this analysis for several different di-jet resonance mass values and consequently estimate the statistical uncertainty on the quark-jet fraction parameter, $\Delta f$, for different resonance couplings and LHC luminosities by considering that $\Delta f$ scales as 
\begin{equation}
\Delta f \sim \frac{\sqrt{1+2 \frac{B}{S}}}{\sqrt{S}}
\end{equation}
with the total number of signal ($S$) and background ($B$) events. $S= \sigma_S L$, $B=\sigma_B L$, where $L$ is the  integrated luminosity, $\sigma_S$ is the di-jet signal cross section (which depends on the resonance mass and coupling), and $\sigma_B$ is the background cross section (which depends on the di-jet invariant mass cut). As in the previous analysis at $M=4$ TeV, we apply the CMS selection cuts in (\ref{eq:cms-cut}) and we restrict to a di-jet invariant mass region $|M_{jj}-M| <  \Gamma/2$, where, conservatively, we take $\Gamma=0.15 M$.

\begin{figure}
\includegraphics[width=0.48\textwidth]{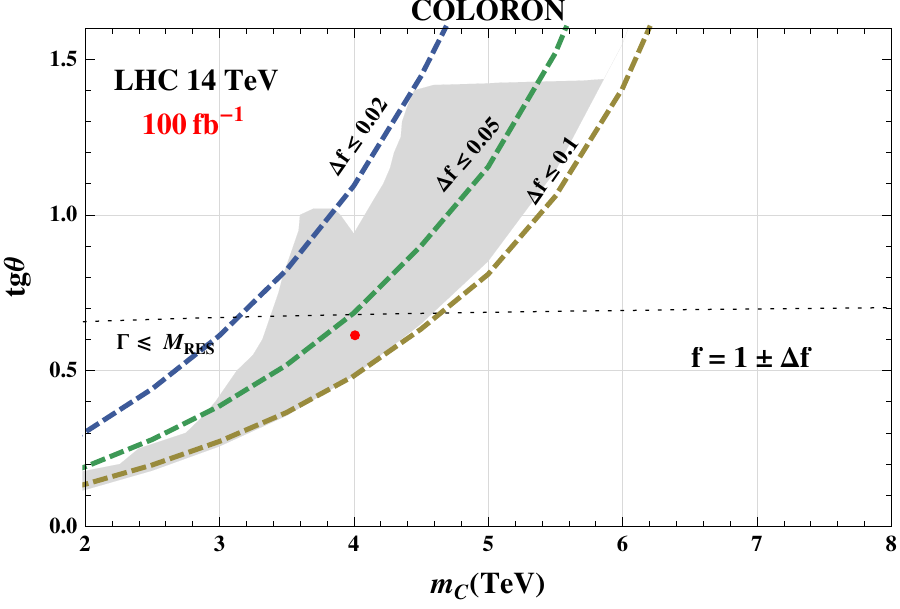}
\includegraphics[width=0.48\textwidth]{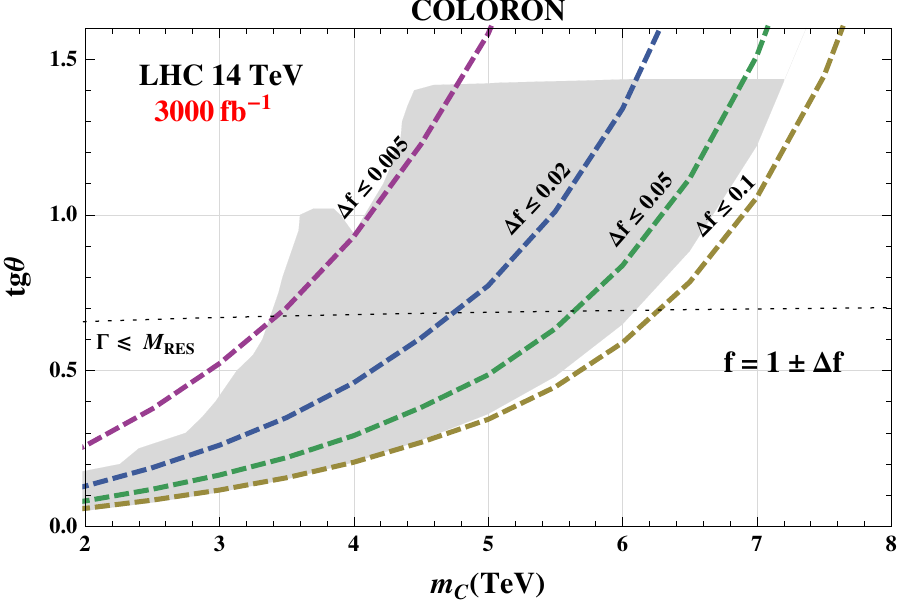}\\
\includegraphics[width=0.48\textwidth]{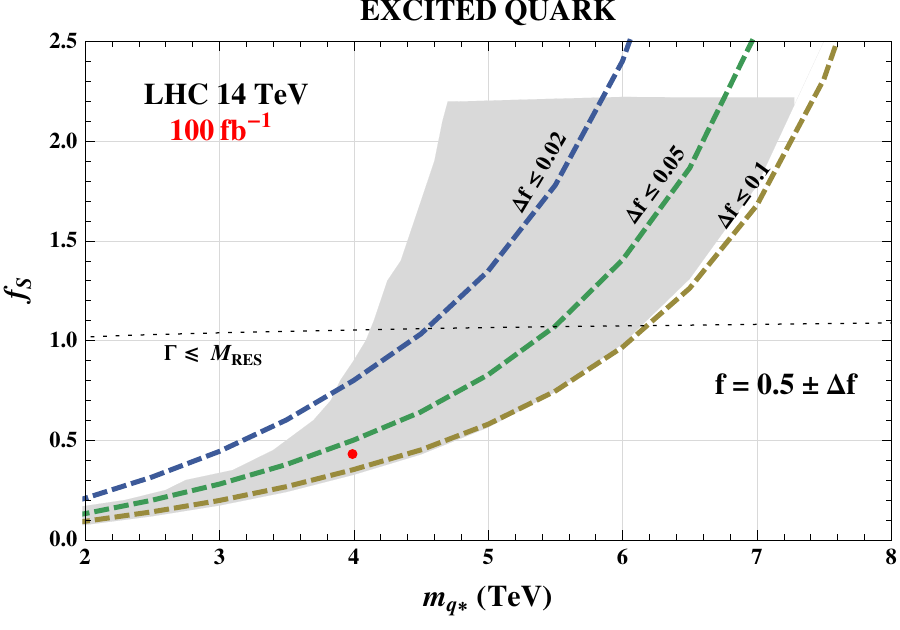}
\includegraphics[width=0.48\textwidth]{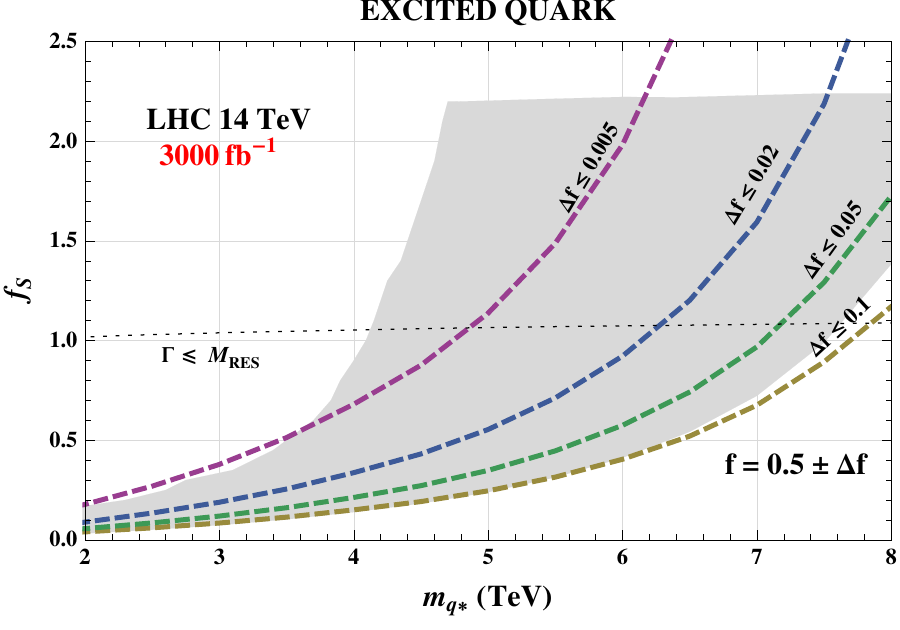}\\
\includegraphics[width=0.48\textwidth]{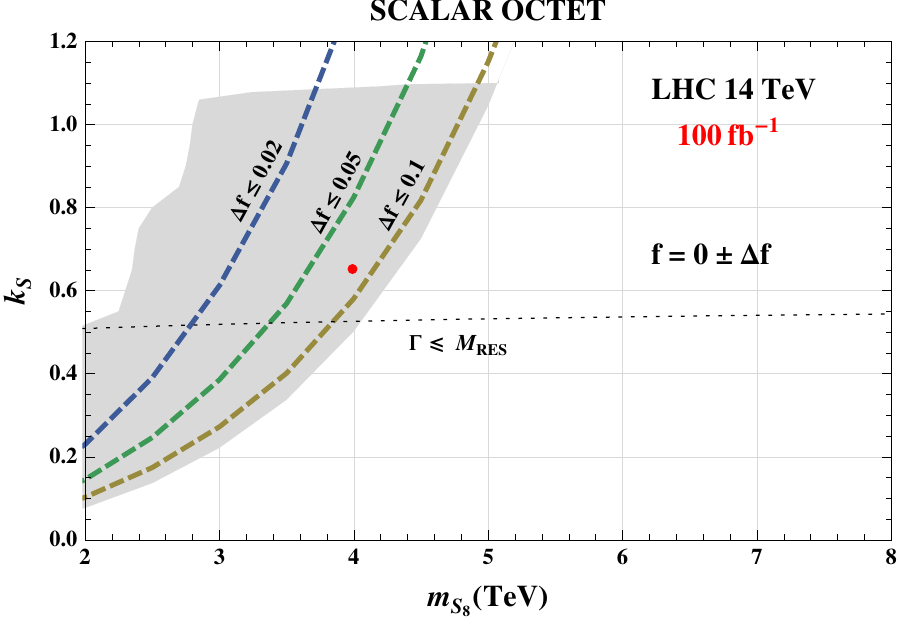}
\includegraphics[width=0.48\textwidth]{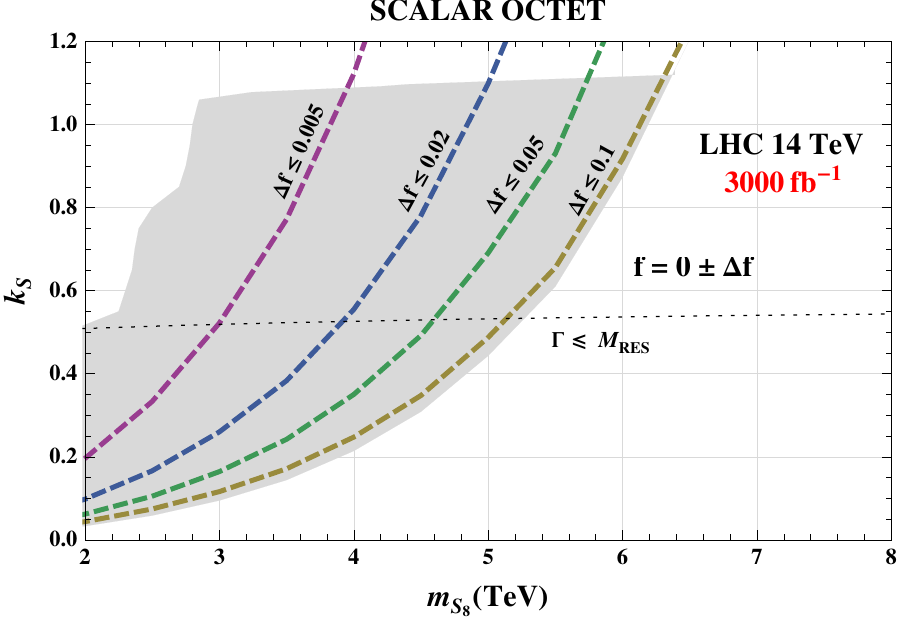}
\caption{\small Contours of constant statistical uncertainty in the quark-jet signal fraction $\Delta f$ (dashed lines) in the mass-coupling parameter space for the three di-jet resonances at different 14 TeV LHC integrated luminosities (Left Plots: 100 fb$^{-1}$, Right Plots: 3000 fb$^{-1}$). The shaded regions show the areas that are allowed by the LHC-8 data, and where we can reach a 5-$\sigma$ discovery of the specific di-jet resonance with the given luminosity. The dotted black line indicates the narrow-width limit; below this line, the color discriminant analysis is not possible but analysis of the JEPs is still valuable. In the case $L=100$ fb$^{-1}$ we also indicate with a red dot the mass-coupling values considered in the benchmark analyses of a $M=4$ TeV di-jet resonance, discussed in the text. Note that the use of JEPs can yield a statistically significant measurement distinguishing between the different types of resonances over the entire parameter space accessible to the LHC.}
\label{fig:Df-contors}
\end{figure}

Through this analysis we can establish the region of masses and couplings where the quark jet-fraction parameter $f$ can be measured sufficiently well to distinguish between colorons, excited quarks, and color-octet scalars. In particular, we obtain contours of constant statistical uncertainty in the signal quark-jet fraction, $\Delta f$, in the parameter space for the three di-jet resonances at different 14 TeV LHC integrated luminosities, as shown in Fig. \ref{fig:Df-contors}. Together with the $\Delta f$ contours we show (in grey) the regions illustrated in Fig. \ref{fig:excl-disc}, that are still allowed by LHC-8 data and where a 5$\sigma$ discovery of the specific di-jet resonance is achievable with the given luminosity. 
In the case $L=100$ fb$^{-1}$ we also indicate with a red dot the mass-coupling values considered in the analysis at $M=4$ TeV. 

The results show that if a 5$\sigma$ discovery of a di-jet resonance occurs  at the 14 TeV LHC, the statistical uncertainty on the corresponding $f$ parameter will be small; we have $\Delta f \leq 0.1$ for all of the three types of resonances in essentially the entire relevant parameter space where we can reach a 5$\sigma$ discovery at the 14 TeV LHC.  Thus, it should be possible to use the analysis of JEPs to distinguish among $gg$, $qg$, and $\bar{q}q$ dijet resonances. 

We must reiterate, however, that our study only examines the statistical significance of the di-jet resonance discrimination through JEPs. We make no attempt to estimate the effects of possible systematic uncertainties on the JEPs, as this will require a detailed detector study and is only likely to be possible with data in hand -- and is therefore beyond the scope of this paper.

\section{Conclusions}\label{sec:conclusions}

We have presented a strategy of distinguishing different possible dijet resonances through the study of the energy profiles of the decay jets. Fig. \ref{fig:Df-contors} summarizes our results for the analysis of the di-jet energy profiles of $\bar{q}q$, $qg$ and $gg$ resonances, including the statistical uncertainties and the effect of background subtraction. We find that the analysis of JEPs can distinguish $gg$, $qg$, and $\bar{q}q$ resonances even after accounting for statistical uncertainties in the signal and the background.

We look forward to exciting results from the upcoming run of the LHC, and the possible discovery of a heavy di-jet resonance.

\section*{Acknowledgments} 
This material is based upon work supported by the National Science Foundation under Grant No. PHY-0854889. We wish to acknowledge the support of the Michigan State University High Performance Computing Center and the Institute for Cyber Enabled Research. PI is supported by Development and Promotion of Science and Technology Talents Project (DPST), Thailand.  RSC and EHS thank the Kobayashi Maskawa Institute for the Origins of Particles and the Universe for hospitality during the SGT15, and also thank the Aspen Center for Physics and the NSF Grant \#1066293 for hospitality during the writing of this Proceedings.

\bibliographystyle{ws-procs975x65}
\bibliography{colvszp}

\end{document}